\documentclass{vgtc}                          

\graphicspath{{figures/}{pictures/}{images/}{./}} 

\usepackage{times}                     

\usepackage{tabu}                      
\usepackage{booktabs}                  
\usepackage{lipsum}                    
\usepackage{mwe}                       

\usepackage{mathptmx}                  
\usepackage{gensymb}
\usepackage{subcaption}
\usepackage{amsmath}
\usepackage{graphicx}
\usepackage{array}
\usepackage[T1]{fontenc}

\onlineid{0}

\vgtccategory{Research}

\vgtcinsertpkg

\title{The Relationship Between Blood Pressure and Self-Reported Stress During Sound-Based VR Relaxation Videos}

 \author{Md Alamin Hossain\thanks{e-mail: alamincsebuet10@gmail.com}\\ %
         Kennesaw State University, USA %
 \and M. Rasel Mahmud\thanks{e-mail: m.raselmahmud1@gmail.com}\\ %
       Kennesaw State University, USA}

 \teaser{
   \centering
   \includegraphics[width=\linewidth]{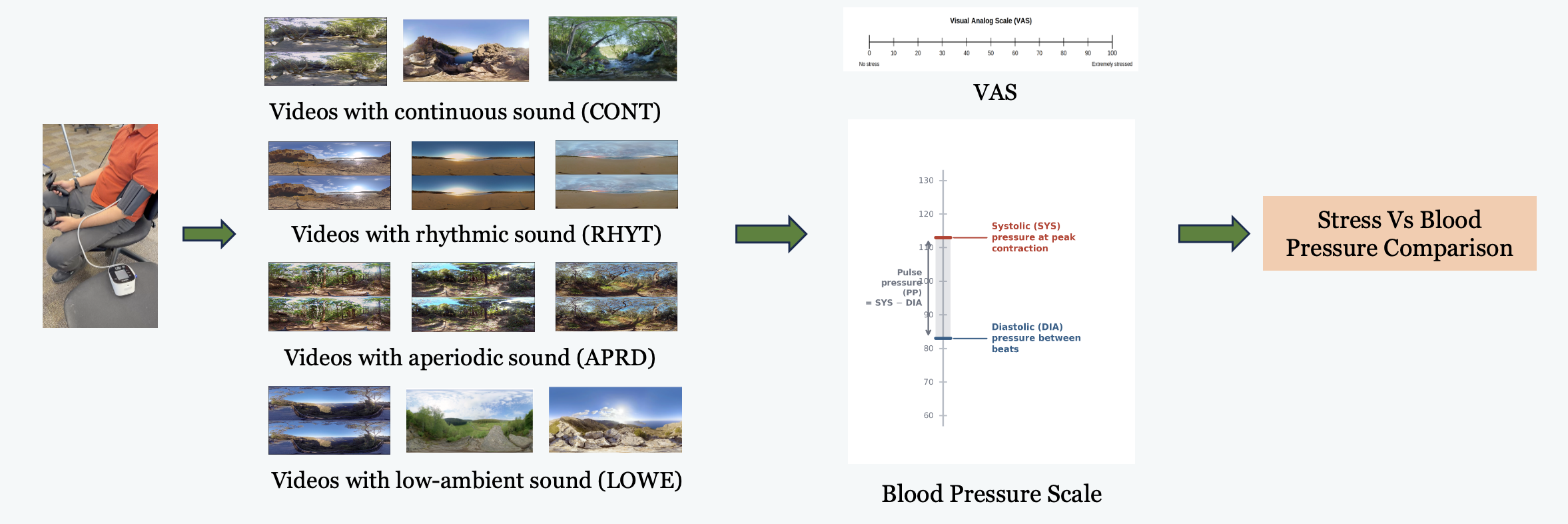}
   \caption{Participants viewed relaxing 360° VR videos across four sound-design categories in a within-subjects design while systolic/diastolic blood pressure and self-reported stress (VAS) were recorded.}
   \label{fig:teaser}
 }

\abstract{
    Virtual reality (VR) relaxation environments are increasingly proposed as accessible tools for stress management, yet the physiological correlates of self-reported stress during VR exposure remain unclear, particularly for blood pressure (BP). We conducted a within-subjects study (N = 18) in which participants viewed relaxing 360$^\circ$ videos, grouped into four sound-design categories, such as videos with rhythmic (RHYT), aperiodic (APRD), continuous (CONT), and low-ambient (LOWE) sound through an HTC Vive Focus Vision headset. Systolic (SYS) and diastolic (DIA) BP were recorded after each category using an Omron 3 Series monitor, and self-reported stress was collected via a 0–100 Visual Analogue Scale (VAS) with fixed intervals. While overall BP did not differ significantly across sound categories, self-reported stress did (Friedman $\chi^2(4) = 12.07, p = .017$), with rhythmic sound eliciting significantly lower stress than aperiodic and continuous sound. Critically, DIA, but not SYS, correlated significantly with self-reported stress across conditions ($r = .31, p = .003$). This suggests that DIA may be a more sensitive physiological indicator of subjective stress than SYS in short VR exposures. We discuss implications for designing and evaluating VR mental health interventions and the value of low-cost physiological sensing alongside self-report.
} 


\begin{document}


\firstsection{Introduction}
\maketitle
Mental health conditions are projected to become the leading global disease burden by 2030, and existing services are struggling to meet demand \cite{chan2024projecting, UNNews2025MentalHealth}. Early intervention, support, and education can meaningfully improve prognosis, and immersive virtual reality (VR) environments have emerged as a scalable, low-risk approach for stress/anxiety reduction, prevention, and support that can be delivered outside traditional clinical settings \cite{hossain2025exploring}. The core mechanism thought to underlie VR's therapeutic effects is immersion, the objective capacity of a system to generate a compelling illusion of presence, or ``being there,'' within the simulated environment \cite{slater2018immersion}. Beyond relaxation applications specifically, VR-based exposure interventions have demonstrated efficacy across a range of anxiety-related disorders \cite{carl2019virtual}. Also, consumer interest in VR relaxation applications has grown rapidly alongside the increasing affordability of consumer-grade headsets \cite{fagernas2021what}. A growing body of work has explored the physiological effects of VR relaxation using markers such as heart rate (HR), heart rate variability (HRV), electrodermal activity (EDA), and blood pressure (BP), electroencephalography (EEG), etc. \cite{robbemond2026physiological}.

Among these physiological markers, BP is particularly attractive for practical deployment in consumer-facing VR mental health tools. It is inexpensive, non-invasive, familiar to both clinicians and lay users, and does not require specialized signal-processing expertise to interpret, unlike HRV frequency-domain measures. At the same time, BP is more susceptible to measurement artifact during movement, and continuous cuff-based measurement is impractical during active VR use. Such constraints likely explain why BP has been used less consistently than cardiac or electrodermal measures in the VR stress literature \cite{robbemond2026physiological}. A further open question, largely unaddressed in prior work, is whether systolic BP (SYS) and diastolic BP (DIA) behave as equivalent indicators of subjective stress or whether they diverge \cite{berton2026psychological, lovallo2003psychophysiological}. This distinction is not merely of academic interest, rather cardiovascular disease remains the leading cause of death and disability worldwide, with blood-pressure-related conditions accounting for a substantial share of this burden \cite{gbd2025global}. This underscores the practical value of clarifying which BP component is most informative for low-cost, consumer-facing stress monitoring.

Secondly, a largely under-explored question concerns the role of sound and vision. Auditory and visual effects have shown great impact in VR study designs \cite{mahmud2022auditory, mahmud2023eyes}. Additionally, sound cues are shown to be effective in reducing mental fatigue \cite{hossain2026cogniaudit} induced by visual or audio feedback applied for resolving accessibility or healthcare issues in VR \cite{mahmud2022auditory2, mahmud2023visual}. A few prior works have investigated the effect of auditory feedback on balance improvement \cite{10670075,cordova2023real,10316460,mahmud2023multimodal,mahmud2022standing}, gait improvement \cite{mahmud2026multimodal, mahmud2025vibrotactile, mahmud2022vibrotactile}, and preventing risk of fall in VR \cite{pavel2025vrfallnet}. However, relatively little work has systematically compared how the acoustic design of a relaxing VR experience shapes physiological and subjective stress outcomes independently of visual content. Prior work on nature-based VR environments has shown that ambient natural sound can support physiological stress recovery \cite{annerstedt2013inducing}, and reviews of VR stress interventions report that most existing work uses generic nature-based content rather than manipulating specific acoustic properties such as rhythmicity or periodicity \cite{robbemond2026physiological}. Understanding which sound characteristics are most calming, and whether this is reflected in BP as well as in self-report, is directly relevant to the design of VR mental health tools intended for real-world deployment.

This paper reports an exploratory within-subjects study examining: 
\begin{enumerate}
    \item whether BP changes across four sound-design categories of relaxing VR video;
    \item whether self-reported stress, measured with a Visual Analogue Scale (VAS), differs across these categories;
    \item whether systolic or diastolic BP is the more sensitive correlate.
\end{enumerate}
The remainder of the paper is organized as follows. Section 2 reviews related work on VR stress interventions, physiological markers, BP reactivity, sound-based relaxation, and VAS measurement. Section 3 describes our method. Section 4 reports our results, and Section 5 discusses their implications and limitations. Section 6 concludes with directions for future work.

\section{Related Work}
\subsection{VR for Stress Reduction and Mental Health}
Several studies have used physiological markers to evaluate VR relaxation interventions. Kim et al. \cite{kim2021effect} compared VR relaxation to conventional biofeedback in a crossover trial with high-stress adults and found that VR relaxation significantly reduced self-reported stress and anxiety, alongside changes in HRV indices, though BP was not the primary focus of that work. Makaroff et al. \cite{makaroff2025novel} evaluated a 30-minute VR relaxation experience with cardiology patients and found a significant reduction in self-reported stress (STAI-S) that only partially converged with physiological measures: HR, BP, and galvanic skin response did not change significantly. This partial dissociation between self-report and physiology directly motivates our focus on which BP component (systolic or diastolic) best tracks subjective stress during VR relaxation.

Beyond biofeedback and cardiology-focused interventions, VR-based mindfulness and meditation programs have also shown promise for stress reduction. Navarro-Haro et al. \cite{navarrohar02017meditation} found that experienced meditators reported significantly greater relaxation and lower sadness after a brief immersive VR mindfulness session, and Modrego-Alarc\'{o}n et al. \cite{modregoalarcon2021efficacy} found that adding VR support to a mindfulness-based program improved treatment adherence among university students, although it did not further reduce stress beyond the non-VR program. Beverly et al. \cite{beverly2022tranquil} similarly found that a brief cinematic VR nature experience reduced subjective stress among frontline healthcare workers during the COVID-19 pandemic, while Cinalioglu et al. \cite{cinalioglu2023effects} note that most VR relaxation research to date has focused on younger adults, motivating extensions to older and more clinically vulnerable populations.

\subsection{Physiological Markers in VR Stress Research}
A recent systematic review of 69 studies using head-mounted-display VR for stress regulation found that HR, HRV, and BP are the most commonly used physiological markers. However, this methodological inconsistency, including small samples, single-session designs, and inconsistent measurement timing, etc., limits comparability across studies \cite{robbemond2026physiological}. The review further notes that most VR relaxation content is nature-based, with comparatively little systematic manipulation of acoustic design, and that many studies treat BP as a single undifferentiated outcome rather than examining SYS and DIA separately. This inconsistency extends to the broader physiological-sensing literature beyond VR specifically: reviews of wearable heart-rate-variability biofeedback report promising but mixed evidence for real-world stress reduction, often limited by low adherence outside controlled laboratory settings \cite{hirten2024remote}, and reviews of wearable electrodermal activity sensing similarly note that most validation work has occurred in lab rather than naturalistic settings \cite{klimek2023wearables}, echoing the methodological concerns raised for VR-specific physiological measurement. Our study addresses both gaps directly.

\subsection{Systolic versus Diastolic Blood Pressure Reactivity}
Blood pressure dysregulation is a central contributor to global cardiovascular disease burden, which remains the leading cause of death and disability worldwide \cite{gbd2025global}, motivating close attention to how psychological stress relates to its two constituent components. Cardiovascular psychophysiology has long distinguished between systolic and diastolic BP reactivity to acute stress, viewing exaggerated cardiovascular reactions to psychological stress as a pathway toward hypertension and cardiovascular disease \cite{berton2026psychological, lovallo2003psychophysiological}. Much of this literature, focused on long-term cardiovascular risk, has emphasized SYS reactivity as a predictor of future hypertension. However, more recent clinical work specifically examining present psychological stress in cardiology outpatients found that self-reported present stress was independently associated with higher DIA, but not with SYS or HR, in multivariable regression models \cite{berton2026psychological}. This recent clinical finding is consistent with the pattern we observe in a VR relaxation context, and suggests that the SYS–DIA distinction in stress reactivity may generalize beyond acute laboratory stressors to lower-intensity, sustained relaxation manipulations.

\subsection{Sound, Soundscapes, and Relaxation}
Annerstedt et al. \cite{annerstedt2013inducing} showed that layering nature sounds onto a VR forest environment supported physiological stress recovery after a stress-induction task, highlighting sound as an active ingredient in VR relaxation rather than a passive backdrop. This is consistent with broader evidence that music and structured sound interventions can reduce physiological and psychological stress markers, including blood pressure, across clinical and non-clinical populations, and that exposure to natural sounds specifically produces small-to-moderate reductions in heart rate, blood pressure, and respiratory rate relative to quiet control conditions \cite{fan2024effect}. A recent scoping review of sound interventions more broadly notes that non-musical sounds, including natural and rhythmic sounds, remain comparatively understudied relative to music despite showing similar promise \cite{saskovets2025effects}, and reviews of simulated natural environments suggest that the degree of technological immersion itself may moderate restorative effects, independent of content \cite{li2023effects}. The perceptual phenomenon of Autonomous Sensory Meridian Response (ASMR), typically triggered by soft, rhythmic, or crisp sounds and associated with subjective relaxation and flow-like states \cite{barratt2015autonomous}, further illustrates these specific acoustic qualities. These studies directly motivated our category-level comparison of rhythmic, aperiodic, continuous, and low-ambient sound.

\subsection{Measuring Subjective Stress with the Visual Analogue Scale}
The Visual Analogue Scale (VAS) is a brief, single-item measure widely used to assess momentary subjective states, including stress and anxiety, and is particularly well suited to repeated within-session measurement where longer multi-item inventories would be impractical. Computerized VAS measures of state anxiety have demonstrated adequate test–retest reliability, convergent validity with the State-Trait Anxiety Inventory, and sensitivity to acute changes in state anxiety following stress induction \cite{abend2014reliability}. This supports our use of VAS ratings taken at two points within each video (30\% and 70\% through playback) as a sensitive, low-burden measure of momentary stress suitable for the fine-grained, repeated-measures design used here.

\section{Methodology}
\subsection{Participants}
Eighteen adults (13 male, 5 female; $M_{age} = 27.8$ years, SD = 2.9, range 24–36) participated in the study. The participants were recruited through online advertisement and the study was approved by the institutional review board (IRB) of the university.

\subsection{Apparatus}
Participants viewed relaxing 360$^\circ$ videos through an HTC Vive Focus Vision head-mounted display. Systolic and diastolic BP and heart rate were recorded using an Omron 3 Series blood pressure monitor attached to the participant's arm. Only SYS and DIA are analyzed in the present paper; HR was also recorded but is not reported here.

\subsection{Stimuli and Sound Categories}
Videos were grouped into four sound-design categories, three videos per category (12 videos in total), plus a no-video baseline period: rhythmic sound (RHYT), aperiodic sound (APRD), continuous sound (CONT), and low-ambient sound (LOWE). All videos were otherwise comparable in visual content so that differences between categories can be more directly attributed to their acoustic design rather than to unrelated visual differences. Each video was around 1 minute and 30 seconds long. All videos were 360$^\circ$ and the participants were able to watch the relaxing scenes through head rotation or through hand-held controllers.

\subsection{Experimental Design}
The study used a within-subjects design with four sound-design categories (No-video, RHYT, APRD, CONT, LOWE) plus a no-video baseline. Video order was counterbalanced across participants to control for order effects.

\begin{figure}
    \centering
    \includegraphics[width=\linewidth]{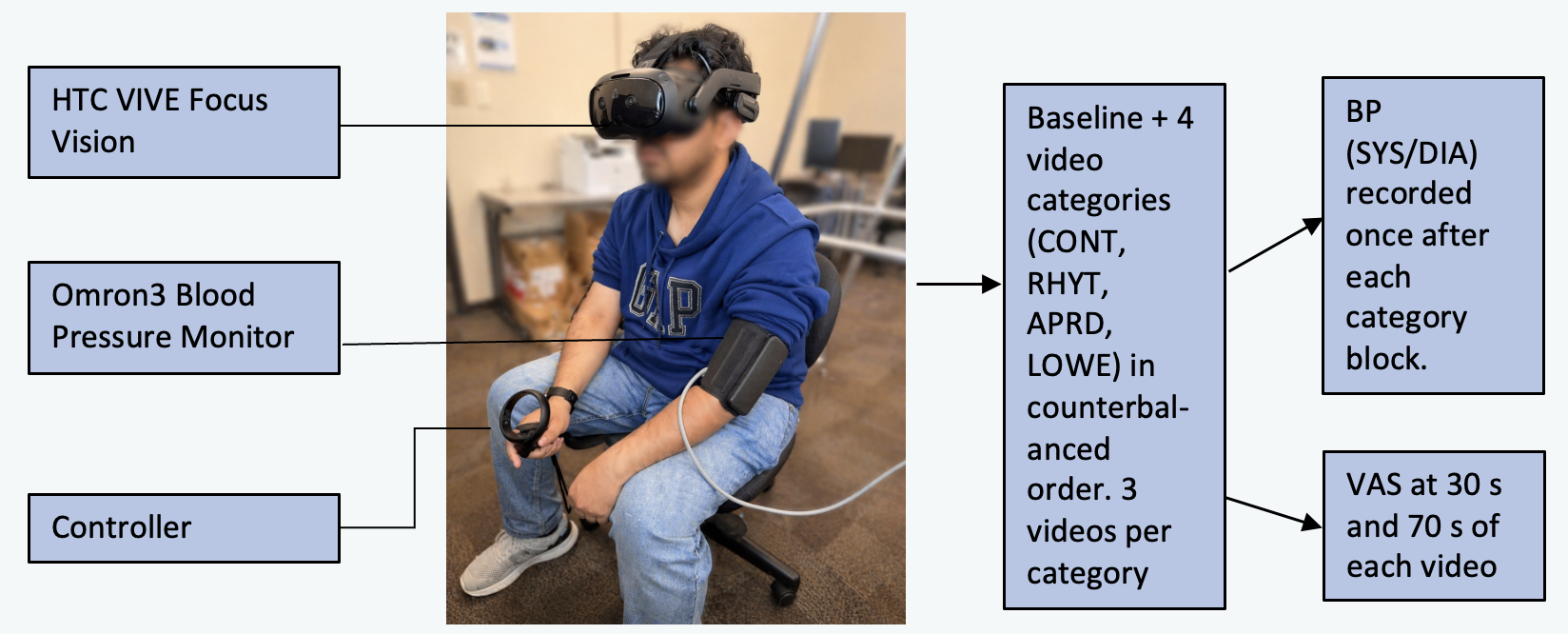} 
    \caption{Experimental setup. Participants viewed three videos per sound category, rating stress via VAS after each video, with one BP reading taken after each category block.}
\label{fig:study_design}
\end{figure}

\subsection{Procedure}
Each participant completed the study while sitting on a chair in relaxed position (Figure \ref{fig:study_design}). After each sound-design category block (three videos), BP was recorded once. During each individual video, participants rated their momentary stress level using a 0–100 VAS (0 = no stress at all, 100 = extremely stressed), completed at approximately 30 seconds and 70 seconds.

\subsection{Measures}
Blood pressure, one SYS and one DIA reading per participant, per sound-design category (Baseline, RHYT, APRD, CONT, LOWE), sitting only, giving 18 participants $\times$ 5 categories = 90 observations. From these, we additionally computed pulse pressure ($PP = SYS - DIA$) for each observation. PP is a BP-derived index more closely associated with arterial stiffness and stroke volume, in contrast to DIA level, which is more closely tied to peripheral vascular resistance; including it lets us test whether any BP-stress relationship is specific to the diastolic level or instead reflects a more general shift in BP amplitude \cite{dart2001pulse, lamia2007contribution}.

Self-reported stress (VAS), two VAS ratings (30 seconds and 70 seconds through playback) were collected per video. We averaged these two ratings per video, then averaged across the three videos within each category, to obtain one VAS score per participant per category, matching the granularity of the BP data.

\subsection{Data Analysis Plan}
We conducted repeated-measures ANOVAs (with Greenhouse–Geisser correction where Mauchly's test indicated a violation of sphericity) to test for differences across the five categories, separately for SYS, DIA, PP, and VAS. Because VAS ratings were positively skewed with many zero values, we additionally report the non-parametric Friedman test as a robustness check, followed by Wilcoxon signed-rank post-hoc comparisons where the omnibus test was significant. We examined the relationship between VAS and each BP measure (SYS, DIA, PP) using Pearson correlations across all participant-by-category observations (N = 90), and repeated this analysis using each participant's change score relative to their own baseline. Finally, we tested whether VAS ratings changed systematically from the 30-second to the 70-second playback point (a within-video habituation check) using a paired t-test across all video trials.

\section{Results}
\subsection{Blood Pressure Across Sound Categories}
\begin{figure}
    \centering
    \includegraphics[width=\linewidth]{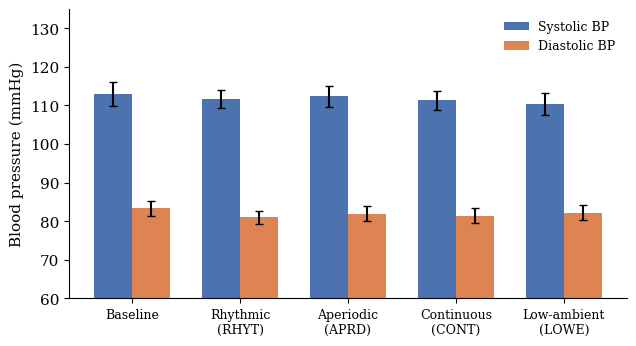} 
    \caption{Mean systolic and diastolic BP by sound category. Error bars show $\pm$ 1 SEM. No significant differences were found across categories.}
\label{fig:bp_res}
\end{figure}

Descriptive statistics for SYS, DIA, pulse pressure ($PP = SYS - DIA$), and VAS are shown in Table \ref{tab:descriptives}. A repeated-measures ANOVA revealed no significant main effect of sound category on SYS, $F(4, 68) = 0.92$, $p = .43$ (Greenhouse--Geisser corrected), or on DIA, $F(4, 68) = 1.12$, $p = .35$. The non-parametric Friedman test confirmed this null result for both SYS, $\chi^2(4) = 0.92$, $p = .92$, and DIA, $\chi^2(4) = 3.62$, $p = .46$. Pairwise comparisons of each category against Baseline were also non-significant for both SYS and DIA (all $p > .10$; Figure \ref{fig:bp_res}). Pulse pressure showed the same pattern: no significant main effect of category, $F(4, 68) = 0.72$, $p = .53$ (Greenhouse--Geisser corrected), confirmed by the Friedman test, $\chi^2(4) = 0.82$, $p = .94$, and no significant Baseline comparisons (all $p > .28$).

\begin{table}[t]
\centering
\caption{Mean (SD) systolic BP, diastolic BP, pulse pressure, and self-reported stress by sound category ($N = 18$). SYS, DIA, and PP in mmHg; VAS on a 0--100 scale.}
\label{tab:descriptives}
\begin{tabular}{lcccc}
\hline
\textbf{Category} & \textbf{SYS} & \textbf{DIA} & \textbf{PP} & \textbf{VAS} \\
\hline
Baseline & 113.1 (13.0) & 83.3 (8.7)  & 29.8 (9.1)  & 14.2 (19.4) \\
RHYT     & 111.7 (10.0) & 81.1 (7.2)  & 30.7 (7.8)  & 11.2 (15.0) \\
APRD     & 112.3 (11.8) & 81.9 (8.1)  & 30.4 (7.5)  & 20.1 (17.2) \\
CONT     & 111.3 (11.0) & 81.4 (8.1)  & 29.9 (9.6)  & 19.0 (19.0) \\
LOWE     & 110.3 (12.3) & 82.2 (8.4)  & 28.2 (10.4) & 16.0 (18.4) \\
\hline
\end{tabular}
\end{table}

\subsection{Self-Reported Stress Across Sound Categories}
\begin{figure}[t]
\centering
\includegraphics[width=\columnwidth]{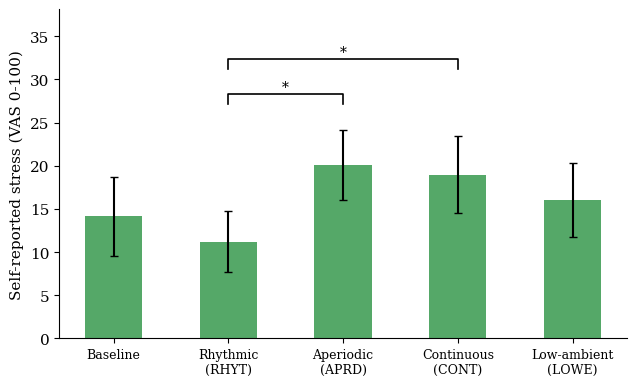}
\caption{Mean self-reported stress (VAS) by sound category. Error bars show $\pm$ 1 SEM. Brackets denote significant post-hoc Wilcoxon comparisons ($^{*}p < .05$).}
\label{fig:vas_by_category}
\end{figure}

Self-reported stress showed a near-significant main effect of category in the repeated-measures ANOVA, $F(4, 68) = 2.42$, $p = .057$ ($p = .090$ Greenhouse--Geisser corrected). The non-parametric Friedman test, better suited to the positively skewed VAS distribution, was significant, $\chi^2(4) = 12.07$, $p = .017$. Descriptively, RHYT elicited the lowest mean stress (11.2), lower than Baseline (14.2), while APRD (20.1) and CONT (19.0) trended higher than Baseline (Figure \ref{fig:vas_by_category}). No individual category-versus-Baseline comparison reached significance (all $p > .10$ by paired $t$-test and Wilcoxon test). Post-hoc Wilcoxon signed-rank comparisons among the four video categories showed that RHYT elicited significantly lower stress than both APRD ($p = .011$) and CONT ($p = .028$), with a similar trend versus LOWE ($p = .070$), indicating that the omnibus effect was driven primarily by rhythmic sound eliciting comparatively low stress relative to the other three sound categories, rather than by any category differing sharply from the no-video Baseline.

\subsection{Habituation Check}
To rule out a simple habituation confound (i.e., stress ratings mechanically declining over the course of each video regardless of condition), we compared VAS ratings taken at 30 seconds and 70 seconds through video playback across all video trials. There was no significant difference, t = 0.42, p = .68 (M30 = 16.65, M70 = 16.50), indicating that momentary stress ratings were stable within a video and that the category differences reported above are unlikely to reflect a simple within-video decay artifact.

\subsection{Relationship Between Blood Pressure and Self-Reported Stress}
\begin{figure}[t]
\centering
\includegraphics[width=\columnwidth]{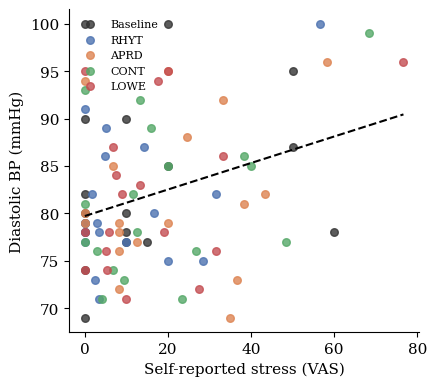}
\caption{Relationship between diastolic BP and self-reported stress (VAS) across all participant-by-category observations, colored by category, with overall linear fit ($r = .31$, $p = .003$).}
\label{fig:dia_vas_scatter}
\end{figure}

Across all 90 participant-by-category observations, DIA correlated significantly and positively with self-reported stress, $r = .31$, $p = .003$, while SYS did not, $r = .16$, $p = .12$ (Figure \ref{fig:dia_vas_scatter}). The same pattern was evident, though weaker, in Spearman rank correlations (DIA: $\rho = .18$, $p = .086$; SYS: $\rho = .02$, $p = .87$), and when computing change scores relative to each participant's own Baseline ($\Delta$DIA vs. $\Delta$VAS: $r = .22$, $p = .065$; $\Delta$SYS vs. $\Delta$VAS: $r = .10$, $p = .40$). Examined within individual categories (Table~2), the SYS--VAS association was significant only for LOWE ($r = .49$, $p = .038$), while the DIA--VAS association showed a similar-sized trend for RHYT ($r = .45$, $p = .060$); neither of these per-category associations survives correction for multiple comparisons given the small per-category sample ($n = 18$), and we treat them as suggestive rather than confirmatory.

We repeated this analysis for pulse pressure ($PP = SYS - DIA$), a BP-derived index more closely tied to arterial stiffness and stroke volume than to peripheral vascular resistance. Unlike DIA, PP showed no significant relationship with self-reported stress, either overall ($r = -.07$, $p = .51$; Spearman $\rho = -.12$, $p = .26$) or within any individual category (Table \ref{tab:correlations}, rightmost column; all $p > .23$), nor when using change scores relative to Baseline ($r = -.07$, $p = .57$). Because PP is arithmetically SYS minus DIA, this null result is informative: it indicates that the DIA--stress association reported above is not simply an artifact of a general shift in overall BP amplitude, but appears specific to the diastolic level itself.

\begin{table}[t]
\centering
\caption{Pearson correlation between self-reported stress (VAS) and SYS, DIA, and pulse pressure (PP), by category.}
\label{tab:correlations}
\begin{tabular}{lccc}
\hline
\textbf{Category} & \textbf{SYS $r$ ($p$)} & \textbf{DIA $r$ ($p$)} & \textbf{PP $r$ ($p$)} \\
\hline
Baseline & -.03 (.90)   & .23 (.37)   & -.26 (.30) \\
RHYT     & .12 (.63)    & .45 (.060)  & -.27 (.29) \\
APRD     & .02 (.95)    & .20 (.44)   & -.19 (.46) \\
CONT     & .25 (.33)    & .40 (.10)   & -.06 (.83) \\
LOWE     & .49 (.038)$^{*}$ & .36 (.14)   & .29 (.24)  \\
\hline
\multicolumn{4}{l}{\footnotesize $^{*}p < .05$} \\
\end{tabular}
\end{table}

\section{Discussion}
\subsection{Diastolic BP as a Candidate Physiological Correlate of Subjective Stress}
The central finding of this study is that diastolic, but not systolic, BP tracked participants' self-reported stress across sound-based VR relaxation conditions. This is a meaningful methodological observation for VR mental health research: much existing work treats BP as a single construct, or defaults to SYS, yet our data suggest DIA may better reflect the sustained autonomic changes associated with subjective stress during a passive relaxation task, while SYS may be more sensitive to acute, momentary, or postural factors less directly tied to relaxation state. This pattern is consistent with a recent multicenter clinical study in cardiology outpatients, which similarly found that present psychological stress was independently associated with higher DIA but not with SYS or HR \cite{berton2026psychological}, suggesting our finding may not be an artifact specific to VR or to our small sample, but may instead reflect a more general dissociation between SYS and DIA as markers of psychological stress that merits further attention in VR mental health research specifically.

Our supplementary pulse pressure analysis reinforces this interpretation. PP did not differ across sound categories and was not associated with self-reported stress in any category, in contrast to DIA. Because PP is arithmetically derived from the same two readings as DIA (PP = SYS $-$ DIA), this rules out a simple explanation in which our DIA finding merely reflects a general widening or narrowing of BP amplitude that happens to load more heavily on DIA than SYS. Instead, the pattern points toward the diastolic level itself, which commonly associated with peripheral vascular resistance and tone. So, DIA is a more relevant physiological channel for subjective stress in this paradigm than PP-related processes such as arterial stiffness or stroke volume. We see this as a useful negative result that sharpens, rather than merely supplements, our main claim.

\subsection{Convergence and Divergence of Physiological and Subjective Measures}
Our results also speak to a broader theme in the VR stress literature: physiological and subjective measures do not always move together. Makaroff et al. \cite{makaroff2025novel} similarly found that self-reported stress improved significantly following VR relaxation while HR, BP, and EDA did not change significantly. This suggests a partial dissociation the authors attributed to the possibility that subjective relaxation can occur without correspondingly large autonomic change, particularly in brief, single-session exposures. Our finding that DIA tracked VAS more closely than SYS suggests that at least part of this apparent dissociation in prior work may reflect an analytic choice. That is to treat BP as one variable, or default to SYS, rather than a true absence of a physiological correlate. We recommend that future VR relaxation studies report SYS and DIA separately, rather than only reporting BP as a single composite or SYS-only outcome, to avoid prematurely concluding that BP is unrelated to subjective stress.

\subsection{Sound Design and Relaxation: Why Rhythm May Matter}
Although overall BP did not differ significantly across sound categories, self-reported stress showed a robust pattern (confirmed by non-parametric testing) in which rhythmic sound (RHYT) was associated with significantly lower stress than aperiodic (APRD) and continuous (CONT) sound. This is consistent with the idea that predictable, rhythmic auditory patterns may support relaxation via entrainment-like mechanisms, in which a regular external rhythm can support a shift toward a calmer physiological and psychological state, echoing prior findings that layering structured natural sound onto VR content supports physiological stress recovery \cite{annerstedt2013inducing}. In contrast, aperiodic or continuously varying sound may require more active auditory processing and thus sustain arousal rather than reduce it. Notably, none of the four categories differed significantly from the no-video Baseline; the effect we observed was specifically that RHYT out-performed the other active sound categories, rather than that any category produced robust relaxation relative to simply sitting quietly. This nuance is important for design: it suggests the choice of sound design among active VR relaxation content matters more, in our data, than the simple presence or absence of an immersive VR experience.

\subsection{Limitations}
There are some limitations that should be considered while interpreting the results of this study. First, the sample size ($N = 18$) is modest for a five-condition repeated-measures design, which limits statistical power, particularly for the per-category correlational analyses reported in Table~\ref{tab:correlations}. Second, due to operating hassle with Omron 3 series device, BP was measured once per category block rather than continuously, so we cannot capture within-block dynamics or align BP as tightly with the finer-grained, per-video VAS data. A continuous or more frequently sampled BP measure would allow closer temporal alignment between physiological and subjective stress signals. Third, our sample was a convenience sample of relatively young, university-affiliated adults (13 male, 5 female; $M_{age} = 27.8$), which may limit generalizability to older adults or clinical populations with elevated baseline stress, anxiety, or other clinical groups. Finally, self-report via VAS, while validated and widely used~\cite{abend2014reliability}, remains subject to demand characteristics and individual differences in stress evaluation, which may not always align with autonomic measures.

\section{Conclusion}
In this study, we examined the relationship between blood pressure and self-reported stress during a sound-based VR relaxation task. While overall BP did not differ significantly across sound categories, diastolic BP was significantly correlated with self-reported stress across conditions, whereas systolic BP and pulse pressure were not. This highlights the value of examining BP components separately rather than as a single composite in VR mental health research. The null pulse pressure result in particular helps rule out a simple BP-amplitude explanation and points toward the diastolic level specifically as the relevant channel. Self-reported stress also showed a robust pattern favoring rhythmic sound design over aperiodic and continuous sound.

Future work should increase sample size and try different BP measuring device to incorporate continuous or more frequently sampled BP measurement (e.g., via continuous cuffless or photoplethysmography-based monitors) to better align physiological and momentary self-report data. Also, extend this sound-design paradigm to clinical populations with elevated baseline stress or anxiety, where the potential benefits as well as risks of VR-based relaxation content are more consequential.

\bibliographystyle{unsrt}

\bibliography{template}
\end{document}